# Evaluation of Time-Series, Regression and Neural Network Models for Solar Forecasting: Part I: One-Hour Horizon

Alireza Inanlougani, T.Agami Reddy and Srinivas Katiamula

School of Computing, Informatics and Decision Sciences, Arizona State University, Tempe, Arizona, USA

School of Sustainable Engineering and The Built Environment, Arizona State University, Tempe, Arizona, USA

Pacific Northwest National Laboratory, Richland, WA, USA

## Abstract

The need to forecast solar irradiation at a specific location over short-time horizons has acquired immense importance. In this paper, we report on analyses results involving statistical and machine learning techniques to predict hourly horizontal solar irradiation at one-hour ahead horizon using data sets from three different cities in the U.S. with different climatic conditions. A simple forecast approach that assumes consecutive days are identical serves as a baseline model against which to compare competing forecast alternatives. One approach is to use seasonal ARIMA models. Surprisingly, such models are found to be poorer than the simple forecast. To account for seasonal variability and capture short-term fluctuations, cloud cover is an obvious variable to consider. Monthly models with cloud cover as regressor were found to outperform the simple forecast model. More sophisticated lagged moving average (LMX) models were also evaluated, and one of the variants, LMX2, identified at monthly time scales, proved to be the best choice.  Finally, the LMX2 model is compared against artificial neural network (ANN) models and the latter proved to be more accurate.  The companion paper will present algorithms and results of how such models can be used for 4-hr rolling horizon and 24-hr ahead forecasting.

**Nomenclature**

| | |
|---|---|
| ANN | Artificial neural network |
| AR | Auto-regressive |
| ARIMA | Auto-regressive integrated moving average |
| ARIMAX | Auto-regressive integrated moving average with exogenous variables |
| CC | Cloud cover |
| CPR | Clean Power Research organization |
| CV-RMSE | Coefficient of variation of the root mean square error |
| I | Horizontal total solar irradiance, $W/m^2$ |
| k | Solar clearness index of atmosphere on hourly time scale |
| LMX | Lagged moving average model with input variables |
| LMX2 | Lagged moving average model with hour of day and month as input variables |
| LR | Linear regression model |
| MAE | Mean absolute error, $W/m^2$ |
| RMSE | Root mean square error, $W/m^2$ |
| $R^2$ | Coefficient of determination |
| SARIMA | Seasonal Auto-regressive Moving Average |
| TMY | Typical Meteorological Year |
| t | Time |
| WU | Weather Underground organization |

## 1. Introduction

Rapid and accelerating growth of solar photovoltaic power installations as a source of renewable energy calls for accurate forecasting of power output. This, in turn, requires forecasting of variables such as solar radiation and ambient temperature, which highly affect power output of a solar system. This capability is also necessary for proper control and power dispatch planning of distributed generation systems. We define the simple forecast approach as one that assumes consecutive days are identical; i.e., the values of the next 24 hours are identical to those of the previous 24 hours for which measured values are available. This would serve as a baseline model against which to compare competing forecast alternatives. The objectives of this paper are to report on the following research questions investigated:

(1) *Radiation forecasting without any exogenous variables such as cloud cover*
What is the accuracy of one-hour ahead radiation forecasts using traditional ARIMA models on a yearly basis? How does this compare with the simple baseline prediction method?

(2) *Radiation forecasting with deterministic cloud cover data*
If cloud cover information is available, what functional form and variables are likely to yield most accurate solar radiation predictions at monthly time scales, and how much improvement can we expect compared to the simple baseline forecast method?

## 2. Literature Review

Following the early work by Goh and Tan (1977), modeling and forecasting for solar radiation has been the focus of numerous studies in the last couple of decades; only the most relevant ones are briefly reviewed here. For a comprehensive review of application of Artificial Intelligence (AI) methods in solar in PV systems one can refer to the work by Mellit et.al (2008a,b) where they concluded that AI methods are of interest in PV systems because they need less computational effort and do not require knowledge of the internal system parameters. Another review paper is that by Inman et.al (2013) who categorized the methods used for developing solar forecasting as regression methods, time series, artificial neural networks (ANN) and other methods, and under each category briefly discussed basic ideas and reviewed the relevant literature.

Kumar and Chendel (2014) also reviewed the work on using ANNs to predict solar radiation. They point out that ANNs predict solar radiation more accurately than the conventional methods, but that their performance is dependent on the input variables used. More recently, Qazi et.al (2015) reviewed the current literature on application of ANN in solar forecasting. They found that prediction performance of neural networks is dependent on input parameters as well as architecture type and training algorithm utilized.

Sfetsos and Coonick (2000) evaluated two different types of ANN; the relative performance of Multi-layer Perceptron (MLP) and recurrent and radial basis function (RBF) and compared their performance to that of conventional methods based on clearness index and found ANNs to be superior. Similarly, Dorvlo et.al (2002) used MLP and RBF to predict the atmospheric clearness index. After training and tested both models using historical data from Oman and comparing their performance, they concluded that models are of similar performance but that RBF needs less computation time. Behrang et.al (2010) investigated MLP and RBF to forecast daily solar radiation considering 6 different combinations of input variables and using data from Dezful, Iran. They drew an interesting conclusion that the optimal set of input variables depends on the ANN model being used.

Reikard (2009) compared the performance of autoregressive integrated moving average (ARIMA), regression, transfer functions, ANN and hybrid models on 6 different data sets and at resolutions of 5 to 60 minutes. He concluded that the choice of the best performing model depends on the resolution of interest.

Benghanem and Mellit (2010) evaluated RBF and MLP models as well as regression models to predict daily global solar radiation. They found that the RBF model outperforms the other two. Rahimkhoob et.al (2013) conducted a comparative study of statistical and ANN models to predict global solar radiation and found ANN to be more accurate. In a similar research, Ahmed et.al (2015) considered MLP, Non-Linear Auto-regressive ANN and an autoregressive (AR) model to forecast solar radiation, and concluded that non-linear autoregressive ANNs have the smallest RMSE.

Lauret et.al (2015) used machine learning techniques along with an AR model to predict solar radiation using historical data from three French islands. They observed that at 4-Hour ahead horizon, machine learning models slightly outperform the Linear AR model but the gap becomes more significant in the case of unstable sky conditions. Koca et.al (2011) used ANNs to predict solar radiation using state and meteorological and concluded that input variables can significantly affect the performance of ANN models. In a similar research, Voyant et.al (2013) used both ARMA and MLP to predict solar radiation under multiple forecasting horizons (which are of great practical application and importance). They found that for 1-hour ahead horizon, the performance of the models are similar; however, for larger forecasting horizons, MLP outperforms ARMA.

For 4-hour forecasting, several researchers, have tried to develop hybrid models by combining the outputs of two or more forecasting methods. These models are supposed to produce better results than the case of using each of the embedded methods individually. Pertinent papers are those by Wang et.al (2015), Monojoly (2017), Wu and Chan (2011), Benmouiza and Cheknane (2013), Voyant et.al (2012), Voyant et.al (2013). Our companion paper also presents results of evaluating forecast methods for 4-hours ahead and 24-hours ahead for three U.S. cities.

## 3. Data Sets Used

We have identified sources of measured climatic data in order to evaluate the predictive accuracy of different forecasting models. The most obvious database is the typical meteorological year (TMY3) data, which is freely available for thousands of locations worldwide. In addition, we have used multi-year historic measured hourly data provided by a general weather services provider, namely Weather Underground (WU) who provides hourly data of irradiation as well as cloud cover. In addition, satellite-measured hourly horizontal solar radiation data for several years was acquired from Clean Power Research (CPR). Historic hourly data of solar radiation from ground-based observations for three cities, i.e. Phoenix, Miami and Chicago which have different seasonal and weather behavior were acquired for the period 2009-2013. For all three cities, the most recent year of data (2013) was chosen as the testing set, while the remaining period prior to that year as the training set.

There were two options on how to select the time series variables: the hourly solar irradiation (I) and the atmospheric clearness index (k). We have evaluated both variables for model fitting. Atmospheric clearness index k has been used by several prior studies on radiation modeling since it detrends the variability in the solar irradiation, namely it removes the effect of deterministic radiation variability due to purely solar geometric variations both diurnally and seasonally. For both the time series variables, we had to adopt the following data cleaning methods, which were determined by elaborate preliminary data analysis evaluations:

(i) **Cleaned data using k**: the feasible range for k was taken to be [0, 0.85]. So, all negative k values are set to zero and all k values greater than 0.85 to 0.85.
(ii) **Cleaned data using I**: the feasible range for I in $W/m^2$ was taken to be [0, 1050]. Thus, all I values lower than 10 were set to zero and all I values greater than 1050 to 1050.

It is worth mentioning that using Statistical and Machine Learning methods to predict solar radiation need a considerable amount of decent quality historical data. The interested reader can refer to Voyant et.al (2017) for an alternative to these methods when such long historical data is not available.

We have used k values to calculate a binary indicator whose value is negative for night hours and positive for daytime hours. This was necessary to correctly calculate the model goodness-of-fit error metrics for both training and validation periods. Furthermore, there were some missing values in the data set. Since time series modeling is involved, such missing data must be rehabilitated. This was done by simply taking the average of the observations before and after the missing value. Note that there are more elaborate ways to deal with missing values in statistical modeling literature. But, since the number of these records was very small (less than 20 points in a year), this simple approach was deemed adequate.

## 4. Seasonal Time Series Models

Seasonal Auto-Regressive Integrated Moving Average Model (SARIMA) predicts the time series values at time t based on the model errors and observations at previous time periods. It also includes differencing and seasonal differencing (Montgomery et al., 2015). The complete SARIMA model is characterized by {(p,d,q)(P,D,Q) T} where p is the order of hourly AR process, and P is the order of the seasonal AR process, d is the order of hourly differencing, and D is the order of seasonal differencing, q is the order of the moving average (MA) process and Q is the order of the MA process for the seasonal part, and T is the number of time period is each cycle. The mathematical form of the SARIMA model is given as:

$$\hat{y}_t = AR(1,1)e_{t-1} + AR(1,2)e_{t-2} + \cdots + AR(1,p)e_{t-p} + MA(1,1)y_{t-1} + MA(1,2)y_{t-2} + \cdots + MA(1,q)y_{t-q} + AR(2,1)e_{t-T} + AR(2,2)e_{t-2T} + \cdots + AR(2,P)e_{t-TP} + MA(2,1)y_{t-T} + MA(2,2)y_{t-2T} + \cdots + MA(2,Q)y_{t-TQ} \quad (1)$$

where $\hat{y}_t$ is the predicted value of the at time t, $y_{t-1}$ is the actual value at time t-1, $e_t$ is the prediction error at time t. Note that the number of parameters in this model is equal to (p+q+P+Q+1). For more detailed discussion on time series, the interested reader can refer to Montgomery et.al (2015), Box et.al (2015) and Madsen (2007).

4.1. Model Training Using I and k

Partial auto-correlation functions were used to determine the maximum possible order of the AR and MA processes. Montgomery et al. (2015) state that a first-order differencing scheme is usually enough for most physical processes. Thus, we considered both regular and seasonal order of differencing to be either 0 or 1 with the seasonality period equal to 24. Depending on the order of the model, these combinations resulted in 216 or 64 different SARIMA models respectively. All these models are fitted to the CPR data and the best models were selected based on the RMSE measure. The model goodness of fit metrics used the coefficient of determination ($R^2$), mean absolute error (MAE), root mean square error (RMSE) and coefficient of variation of the RMSE (CVRMSE). These metrics are well known and defined in any statistical textbook.

Note that the MAE, RMSE and CVRMSE values reported against the variable k are not for k but for the prediction of hourly irradiation I which can be deduced from k. This way of reporting the error metrics allows the relative performance of different models to be compared directly. From Table 1, we observe that for all three cities, models based on only last year's data (2012) have better error metrics for the training data. More importantly we concluded that, using I as the time series variable provides models with better internal error metrics than using k.

4.2. Model Validation

The optimal models identified above were applied to the testing data set (hourly values for year 2013) and error metrics were calculated for each of the models developed with different training data sets. These metrics along with those calculated from the simple baseline forecast method are assembled in Table 2. The MAE and RMSE are in W/m$^2$ while the CVRMSE is a fraction.

The following important conclusions can be drawn from Table 2:

i) For all three cities, the prediction accuracy improves when using I instead of k as the time series variable.

ii) Comparing Table 2 with Table 1, we observe that good internal error metrics do not necessarily imply good prediction performance on the testing set, and this degradation in performance could be quite large. For example, in Table 1 for Chicago, the CVRMSE value for the model for I in 2012 is 0.27, which degrades to 0.68 for testing (from Table 2). This suggests that some of the models identified are not very robust.

iii) Most importantly, the simple forecast method provides predictions that are almost as good as those from SARIMA models. This indicates that there is no benefit in adopting a SARIMA modeling approach to serve as a single annual solar irradiation forecast model.

### 4.3. Residual Analysis

To investigate the rather unexpected observation stated in (iii) above, residual and time series plots for several days and several months have been generated and studied. We found that variability in successive days during different months to be the main cause for the deficient performance of the annual SARIMA models. As seen from the illustrative plot of Figure 1(a), the radiation during the last three days is very much different from the other relatively clear days, and such weather fluctuations cause the annual SARIMA model to perform poorly (see the residual plots of Fig 1b). We conclude that: (i) irradiation is season-dependent which is different for various locations, and so monthly models are probably better at capturing long-term behavior than a single annual model, and (ii) some variable which can capture short-term weather variability ought to be included in the model. An obvious surrogate is the hourly cloud cover variable.

## 5. Linear Regression (LR) Models

As noted from Figure 1, radiation patterns on two consecutive days could be quite different and this calls for adding cloud cover to be included as an input variable to the solar forecasting model. It is worth mentioning that cloud cover values are point records whereas radiation values are averaged values. So, for data consistency we should consider the difference in averaged could cover as the input variable. We have evaluated three different linear regression models, namely, the LR, LMX and LMX2 discussed below.

### 5.1. LR Model Using CPR Data

This LR model assumes that radiation at period t depends on the known radiation value at time t-24 and the difference in hourly cloud cover:

$$\hat{I}_t = \beta \cdot I_{t-24} + \alpha(cc_t - cc_{t-24}) \qquad (2)$$

where cc is the average cloud cover and $I_t$ is the radiation at time t. One interesting feature of this LR model form is that it can be used to make several predictions at consecutive hours during the forecast horizon without being corrupted from error accumulation. Note however, that during actual implementation of these modeling equations, the cloud cover variable will also have to be forecasted.

The monitored hourly data for the same three cities selected were used for model training. Both a single annual model and individual monthly models were identified using the training data set (2012 data) and evaluated on the test set (2013 data), and the corresponding model error metrics are assembled in Table 3. The baseline model results are also shown as "Simple Forecast". We note the following:

(i) In all cases, using monthly LR models result in smaller RMSE values compared to those from the simple forecast method. However, the simple forecast method results in lower MAE values compared to the linear regression model, since the Least Squares method used to estimate the model parameters minimizes RMSE and not MAE.

(ii) The improvement in using monthly LR models is not as significant as we had hoped. The solar radiation data from CPR dataset comes from ground based measurements while the cloud cover was derived from satellite-data. This inconsistency could be the reason why the improvement is not significant.

### 5.2. LR Model Using TMY3 Data

Due to the inconsistency in CPR data as noted above, linear regression analysis was redone but using TMY3 dataset where both cloud cover and solar radiation are measured on the ground. Note that the TMY3 data for the year is made up of monthly data from different years with each month reflective of the long-term average behavior of that month. Hence, fitting an annual model to this data maybe misleading, and therefore only monthly LR models have been fitted. For each month, the data for the last week is left out and is taken to be the testing data set. The results are summarized in Table 4 from which we draw the following conclusions:

(i) For all months and cities, the LR model outperforms the simple forecast model. However, the improvement is different for different months based on the seasonal characteristic of the specific location. More specifically, note that for June in Phoenix, the simple forecast method and the LR model have very similar error metrics, as are those for Miami for March and June, all of which are clear months. However, for Chicago, a significant improvement in the LR model is noticed for all the months. These findings are consistent with our intuition that for those months which are clear and sunny or uniformly cloudy, each day is very similar to the next and so a simple forecast model would suffice. On the other hand, for the months exhibiting wide solar variability during consecutive days, the use of the LR model greatly improves the accuracy of the solar radiation forecasts.

(ii) The improvement between the two modeling approaches is much larger than those found in the previous section where the cloud cover data was satellite-derived while solar radiation values were ground observations. This emphasizes the fact that we should be careful in the source of data sets being selected for regression model training.

The next section discusses further improvement in the model structure involving modification to the regressor terms.

**6. LMX Models**
In this section, we introduce a forecasting approach that can be viewed as an extension of the traditional ARIMA models which involves introducing input variables along with their lagged terms. More specifically, the aim is to capture both the moving average (MA) structure in the data and the effect of cloud cover as an input variable. These models are basically linear regression models with radiation lags (both seasonal and non-seasonal) and the difference in cloud cover as input variables. We expect to obtain the best possible performance by applying such types of models.

6.1. Model Mathematical Forms

The functional form of LMX model identified for individual months is given by:

$$\hat{I}_t = \sum_{i=1}^{2} \beta_i I_{t-i} + \beta_5 I_{t-24} + \sum_{i=1}^{2} \alpha_i (\widehat{cc}_t - cc_{t-i}) + \alpha_5 (\widehat{cc}_t - cc_{t-24}) \quad (3)$$

Another version of LMX model called Lagged De-Trended Moving Average with input variables (LDMX) can be framed as:

$$\hat{I}_t = \sum_{i=1}^{2} \beta_i (I_{t-i} - I_{t-i-1}) + \beta_5 I_{t-24} + \sum_{i=1}^{2} \alpha_i (\widehat{cc}_t - cc_{t-i}) + \alpha_5 (\widehat{cc}_t - cc_{t-24}) \quad (4)$$

where $\hat{I}_t$ is the predicted radiation at time t, $I_{t-i}$ is the measured radiation and time lag i, $\widehat{cc}_t$ is the predicted cloud cover at time t and $cc_t$ is the measured cloud cover at time t.

A third variant is one that considers "time" or hour of day as one of the input variables and month as a categorical variable. Such a model, called LMX2 model, would assume the following form:

$$\hat{I}_t = \sum_{i=1}^{2} \beta_i I_{t-i} + \beta_3 I_{t-24} + \sum_{i=1}^{2} \alpha_i cc_{t-i} + \alpha_3 cc_{t-24} + \alpha_4 \widehat{cc}_t + \alpha_5 Time + \alpha_6 Month \quad (5)$$

The variable "Time" is a categorical variable in the range [1,24] representing the hour of day index, and the variable "Month" is also a categorical variable for the month of the year index. Note that the LMX2 model is a single function which can be used for irradiation prediction for the entire year, while LMX models should be identified for each month individually.

Finally, we can also assume a monthly variant of LMX2 model as follows:

$$\widehat{I_t} = \sum_{i=1}^{2} \beta_i I_{t-i} + \beta_3 I_{t-24} + \sum_{i=1}^{2} \alpha_i cc_{t-i} + \alpha_3\, cc_{t-24} + \alpha_4 \widehat{cc_t} + \alpha_5 Time \quad (6)$$

Note that this approach requires 12 different models which any of them would be more accurate than the single annual LMX2 model.

6.2. LMX model Validation

We have evaluated four different forms of LMX models (eq.3) distinguished by the number of hourly lags (namely 1, 2, 3 and 4 lags). We have also considered two different modeling scenarios. Under the first scenario called "model using all data", we have fit the model to all data points over the entire 24 hours. Under the second modeling scenario, we filtered the data points and used only the daytime values to fit the model. It is worth mentioning that the models are trained using data from the first three weeks of each month and are tested on the data for the last week. Due to space limitation, the results for only the optimal models are shown in Table 5. However, the conclusions drawn from the full set of results are stated below.

i) For most of the months and cities of interest, the model using daytime data outperforms the model using all data. The only exception is June in Phoenix for which the metrics are close.

ii) Both modeling scenarios provide us with a much better predictive model than one based on the simple forecast method. These models outperform simple forecast because they are able to capture the time series behavior of the radiation values and more importantly, include the expected change in cloud cover.

iii) The models of order either 1 or 2 seem to be the best fits to the data. Therefore, we recommend that both models ought to be fit to the data for each month, and the better model selected for implementation.

iv) The gap in performance prediction between the proposed model and simple forecast method is not the same for different months. For example, in Phoenix, the difference is much larger in October and December than it is in June and August. This could be due to consistent and repetitive day to day behavior patterns during June and August. On the other hand, the change in radiation from one day to another is significant in October and December.

Figure 2 allows us to compare the residual values for the simple forecast method and LMX model for three days of the test data set, for Chicago and in March. It can be observed that the simple forecast method is not very robust and suffers from the drawback that it could lead to enormous prediction errors during certain time periods attributed to substantial changes in cloud cover in consecutive days.

6.3. LDMX Model Evaluation

This section presents the results of our evaluation of LDMX models (eq.4). We evaluated four different model forms distinguished by the number of lags (namely 1, 2, 3 and 4 lags). We also consider two different modeling scenarios. Under the first scenario called "model using all data", we fit the model to all data points. Under the second modeling scenario, we filter the data points and just use daytime values to fit the model. Due to space limitation, the results for only the optimal models are assembled in Table 6. However, the conclusions drawn from the full set of results are as follows:

i) For all the months and cities of interest, the model using daytime data outperforms the model using all data.

ii) The model using Daytime data provides us with a much better predictive model than one based on the simple forecast method. We should note that the simple forecast method only considers the radiation value one day previously, while the proposed LDMX model is better able to capture the time series behavior of the radiation values and more importantly, include the expected change in cloud cover. This is probably the reason for such a substantial improvement in prediction accuracy.

iii) The models of order either 1 or 2 seem to be the best fits to the data. Therefore, we recommend that both models ought to be fit to the data for each month, and the better model selected for implementation. Note that compared to Table 5 a model of order 1 suffices for most of the cases because de-trending causes the second lag to get involved in the model indirectly.

iv) The gap in performance prediction between the proposed model and simple forecast method is not the same for different months. For example, in Phoenix, the difference is much larger in October and December than it is in June and August. This could be due to consistent and repetitive day to day behavior patterns during June and August. On the other hand, the change in radiation from one day to another is significant in October and December.

v) Note that there is a significant difference in MAE and RMSE values which implies the presence of some data points with extremely large residuals.

Table 7 below is a summary of Tables 5 and 6 showing the average of optimal CVRMSE value over different months for different model forms and modeling scenarios. The most significant conclusion is that the LMX model is more accurate than the LDMX model for all three cities considered. The improvement is CVRMSE is 11% to 23%. The LMX model is also about twice more accurate than the simple forecast.

6.4. Model Comparison Summary

Performance differences of the different models can be clearly noticed in Figure 3 specific to Phoenix and Chicago. These scatter plots allow us to visually gauge the differences in CVRMSE values of the simple forecast (SF), and the two linear regression variants (LR, LMX) for the best and worst months for the testing data set. The extent to which the error statistics are superior for the LMX model compared to the other two models is clearly seen.

We conclude that:

i) For Phoenix, the performance of the LMX model and that of the LR model are close. However, this is certainly not the case for Chicago. This implies that for Chicago, the moving average structure of the radiation data plays a much greater role than it does for Phoenix.

ii) More importantly, it can be observed that the gap between the best and the worst performing months of the LMX method is not as large as compared to those of the other methods; this is certainly a big advantage of this model type.

iii) Finally, we observe that the model prediction performance is different for different months. Also, its pattern is different for different cities based on the local weather condition. Thus, identifying individual monthly models, though more tedious is the best modeling approach.

6.5. LMX2 Model

The LMX2 model (eq. 5 and 6) has been evaluated against the LMX model. TMY 3 data is assumed for the same three cities and 10-fold cross-validation instead of hold-out method model identification is used to obtain a more reliable estimate of the error metrics. The error estimates for LMX and the simple forecast method are averaged over all months of the year and assembled in the Table 8. On the other hand, LMX2 is an annual model and no averaging is needed.

From this table, the following conclusions can be drawn:

i) The simple forecast is significantly poorer than the other two models.

ii) LMX2 model outperforms the LMX model for Phoenix by a large amount and this applies to Chicago as well. This implies that inclusion of non-seasonal lags of cloud cover as well as "Time" as input variables have contributed to better prediction performance of the LMX2 model

iii) However, this is not the case for Miami even though the LMX model is only slightly better. This is counter-intuitive, but may be since the LMX model requires individual monthly models to be fit which can compensate for the loss in prediction performance caused by not having "Time" as an input variable in the model. Thus, for Miami the monthly variability of solar radiation plays a greater role than does daily variability.

6.6. Monthly LMX2 Models

The annual LMX2 models described in section 6.5 in which the month index is assumed to be an input variable, a single model developed using the entire annual data set. This is a significant advantage for subsequent deployment. One the other hand, a single model could result in poorer prediction accuracy since the radiation dynamics may vary seasonally in a location. To evaluate whether, and to what extent, radiation forecast accuracy would increase when individual monthly LMX2 models are identified, the LMX2-M models for 1-Hour ahead predictions for the same 6 months using TMY3 data are tested using a 10-fold cross validation method. To compare the performance of these models to that of annual LMX2 models we have averaged the error metrics over months, and the results are shown in Table 9. We conclude that the average of RMSE and CVRMSE values for monthly LMX2 model is lower than that for annual LMX2 model.

7. ANN Model

Most of the published papers found ANN to the most accurate model for solar forecasting. An ANN model is composed of an input layer, one or more hidden layers and one output layer. In each layer, there are several nodes each with a bias parameter and a set of weight parameters that are multiplied by the inputs to that node. Finally, a transfer function is applied to this weighted sum to produce the node output. Note that in the input layer, nodes do not have transfer function and their output is simply the variable values. See Bishop (1994), Haykin (1994), Hagan (1996) and Fausett (1994) for more details on neural networks.

7.1. ANN Architecture

A commercial package was used to fit and evaluate the ANN model. ANN architecture, is basically defined by network parameters namely, the number of hidden layer, the number of hidden units in each layer, transfer function and gradient descent parameters i.e. learning rate, validation set etc. Standard version of back propagation algorithm was employed to train the neural networks even though other forms can be used (Neelmegam and Amitham, 2016).

The input variables are the same as those used with the LMX2 model, one hidden layer composed of 11 nodes was assumed and the sigmoid transfer function was used for all hidden units:

$$s(t) = \frac{1}{1+\exp(-t)} \quad (7)$$

We have first fitted a neural network with default settings and then parameters were tuned based on a trial and error approach to yield a better model. Table 10 shows the default and optimal values of the parameters found along with the corresponding error metrics.

The significant improvement in the error metrics and the numerical differences between the default parameter settings and the tuned values reported in Table 7, emphasize the need for proper ANN model identification.

### 7.2. ANN Model Validation

Using the network structure discussed above, we have fitted neural networks to TMY3 and CPR data for the same three cities the 10-fold cross-validation method was adopted which does not require separate training and testing data sets. The forecast model error metrics for models identified from different data sets are assembled in Table 11 from which we draw the following conclusions:

i) Prediction performance of the models trained using the whole CPR dataset is much better than those obtained using just the most recent dataset. This is true for both neural network and linear regression model and for all three cities of interest.

ii) Comparing the prediction performance of the models trained using CPR dataset and those trained using TMY3 data set, we note that using CPR dataset provides us with better predictive models. This holds true for all three cities studied.

iii) Neural network models outperform the linear regression for all datasets and cities of interest (supporting numerous published studies).

iv) The prediction performance of the models is consistent for all cities. Also, the trend is similar for the neural network and the linear regression model, i.e. if city A neural network has better prediction performance than that for city B, the same would be true for the linear regression model as well.

### 8. Summary and Conclusions

The annual SARIMA models were used to investigate the choice of the random variable to be forecasted. Atmospheric clearness Index (k) and solar radiation (I) considered as options. The analysis was done with hourly data for several years from three locations. It was found that using random variable I along with the most recent year's data would lead to the most accurate model. However, this modeling approach was found to be no better than the baseline strategy where each day is assumed to be similar to the previous day. Studying the residual plots let us to conclude that the modeling should be done on a monthly time scale and that an input variable was needed to capture the diurnal and seasonal weather patterns in the data. The most obvious variable is the cloud cover, and this variable was assumed to be known in all subsequent analyses.

Next, we evaluated linear regression (LR) models which included hourly observed cloud cover as an input variable. Such models need to be identified for each month of the year. Using LR instead of Baseline model the decrease in RMSE was in the range 4% to 72% and its average was 15%.

In addition, the other variant, called LMX model, was evaluated which allows us to include the moving average structure of both cloud cover and previous radiation values. Using these models instead of LR model, the decrease in RMSE was in the range 10% to 71% and its average was 39%. Also, comparing

LMX model to the Baseline model, the corresponding range was 13% to 79% with its average being 48%. Further improvement in model forecasting accuracy was achieved by adding the month index (1 to 12) and time index (1 to 24) as input variables along with two lagged terms for radiation and cloud cover and developing single annual model called LMX2. Further improvement to LMX2 model can be achieved by fitting individual monthly models which do not include month as one of their input variables.

Using monthly LMX2 instead of LMX2 model, could help us to decrease RMSE by 18% on average. This was 13% for switching from LMX to LMX2. It is worth mentioning that monthly LMX2 models could help us to decrease RMSE by 58% as compared to the Baseline model.

We concluded the companion paper by comparing ANN to LMX2 models. We found ANNs to be better (consistent with published literature) for the three cities assessed. The ANN models could help to decrease RMSE by 5% on average compared to that using monthly LMX2 models to be more specific. However, training and deployment of neural network requires some expertise, and, moreover, the model terms and coefficients are opaque and not interpretable, which could limit their widespread application in field installations.

The companion paper will present algorithms and results of how such models can be used for 4 and 24 hours ahead forecasting.

## Acknowledgements

This work was supported by the Buildings Technologies Office of the U.S. Department of Energy's Office of Energy Efficiency and Renewable Energy [under Contract DE-AC05-76RL01830 thru Pacific Northwest National Laboratory]. The authors thank Mr. Joseph Hagerman, Technology Development Managers for his guidance and strong support of this work. We acknowledge George Hernandez from PNNL for his technical guidance. We also thank Joe Huang for supplying us with much of the solar radiation data used in our analysis.

**Tables**

Table 1. Error Metrics for Optimal SARIMA Models During Training

| City | Variable | Training Data | Optimal Model | MAE | RMSE | CVRMSE |
|---|---|---|---|---|---|---|
| Phoenix | I | 2009-12 | (1,0,0) (0,1,1) 24 | 16 | 43 | 0.18 |
|  | I | 2012 | (2,0,1) (0,1,1) 24 | 12 | 30 | 0.13 |
|  | k | 2009-12 | (1,0,0) (1,1,0) 24 | 42 | 82 | 0.34 |
|  | k | 2012 | (1,0,1) (0,1,1) 24 | 188 | 338 | 1.39 |
| Chicago | I | 2009-12 | (2,0,0)(0,1,1) 24 | 28 | 55 | 0.32 |
|  | I | 2012 | (2,0,0)(1,1,1) 24 | 23 | 47 | 0.27 |
|  | k | 2009-12 | (0,1,0)(1,1,1) 24 | 218 | 61 | 0.35 |
|  | k | 2012 | (0,1,0)(0,1,1,) 24 | 22 | 48 | 0.28 |
| Miami | I | 2009-12 | (2,0,0)(1,1,1) 24 | 35 | 73 | 0.36 |
|  | I | 2012 | (2,0,0)(1,1,1) 24 | 30 | 62 | 0.3 |
|  | k | 2009-12 | (1,0,0)(0,1,1) 24 | 36 | 76 | 0.37 |
|  | k | 2012 | (1,0,0)(0,1,1) 24 | 32 | 66 | 0.33 |

Table 2. Error Metrics for SARIMA Model Validation for 2013 Based on Models Identified from Various Training Data Sets

| City | Variable | Training Data | MAE | RMSE | CVRMSE |
|---|---|---|---|---|---|
| Phoenix | I | 2009-12 | 26 | 59 | 0.25 |
|  | I | 2012 | 150 | 253 | 1.14 |
|  | k | 2009-12 | 201 | 325 | 1.36 |
|  | k | 2012 | 100 | 161 | 0.67 |
|  | Simple Forecast | 2012 | 28 | 90 | 0.37 |
| Chicago | I | 2009-12 | 60 | 115 | 0.72 |
|  | I | 2012 | 56 | 109 | 0.68 |
|  | k | 2009-12 | 83 | 157 | 0.99 |
|  | k | 2012 | 72 | 136 | 0.85 |
|  | Simple Forecast | 2012 | 57 | 129 | 0.75 |
| Miami | I | 2009-12 | 49 | 98 | 0.49 |
|  | I | 2012 | 65 | 124 | 0.62 |
|  | k | 2009-12 | 119 | 235 | 1.17 |
|  | k | 2012 | 66 | 125 | 0.62 |
|  | Simple Forecast | 2012 | 57 | 130 | 0.64 |

Table 3. LR Model Validation Error Metrics Using CPR Data

| City | Data | Linear Regression | | | Simple Forecast | | |
|---|---|---|---|---|---|---|---|
| | | MAE | RMSE | CVRMSE | MAE | RMSE | CVRMSE |
| Phoenix | 2013 Whole Year | 43 | 90 | 0.38 | 33 | 96 | 0.4 |
| | 2013 March | 64 | 111 | 0.47 | 50 | 123 | 0.52 |
| | 2013 June | 19 | 44 | 0.13 | 15 | 49 | 0.14 |
| | 2013 December | 29 | 63 | 0.45 | 22 | 66 | 0.48 |
| Chicago | 2013 Whole Year | 77 | 128 | 0.8 | 63 | 141 | 0.88 |
| | 2013 March | 84 | 130 | 0.86 | 64 | 142 | 0.98 |
| | 2013 June | 120 | 185 | 0.76 | 111 | 204 | 0.86 |
| | 2013 December | 41 | 71 | 1.06 | 33 | 79 | 1.18 |
| Miami | 2013 Whole Year | 96 | 148 | 0.74 | 90 | 155 | 0.77 |
| | 2013 March | 100 | 152 | 0.69 | 95.1 | 161 | 0.72 |
| | 2013 June | 108 | 160 | 0.68 | 100 | 167 | 0.71 |
| | 2013 December | 66 | 103 | 0.76 | 62 | 107 | 0.78 |

Table 4: LR Model Validation Error Metrics Using TMY Data

| City | Month | Linear Regression | | | Simple Forecast | | |
|---|---|---|---|---|---|---|---|
| | | MAE | RMSE | CVRMSE | MAE | RMSE | CVRMSE |
| Phoenix | January | 44 | 86 | 0.6 | 38 | 96 | 0.67 |
| | March | 140 | 203 | 0.41 | 204 | 296 | 0.6 |
| | June | 50 | 98 | 0.32 | 51 | 105 | 0.34 |
| | August | 43 | 78 | 0.26 | 32 | 82 | 0.27 |
| | October | 31 | 57 | 0.3 | 25 | 70 | 0.37 |
| | December | 124 | 160 | 0.51 | 123 | 194 | 0.61 |
| Chicago | January | 52 | 80 | 1.16 | 40 | 87 | 1.27 |
| | March | 61 | 109 | 1.08 | 65 | 137 | 1.26 |
| | June | 118 | 162 | 0.64 | 105 | 176 | 0.7 |
| | August | 105 | 142 | 0.64 | 93 | 158 | 0.71 |
| | October | 77 | 109 | 0.95 | 66 | 123 | 1.08 |
| | December | 48 | 77 | 1.14 | 43 | 86 | 1.28 |
| Mimai | January | 52 | 97 | 0.61 | 54 | 117 | 0.74 |
| | March | 56 | 86 | 0.35 | 47 | 89 | 0.36 |
| | June | 72 | 114 | 0.46 | 57 | 118 | 0.48 |
| | August | 77 | 136 | 0.6 | 71 | 154 | 0.69 |

| | October | | 53 | 96 | 0.59 | | 50 | 352 | 2.19 |
| | December | | 60 | 99 | 0.81 | | 50 | 106 | 0.86 |

Table 5. LMX Models Validation Error Metrics Using TMY Data

| City | Month | Order | All Data | | | Daytime Data | | | Simple Forecast | | |
|---|---|---|---|---|---|---|---|---|---|---|---|
| | | | MAE | RMSE | CVRMSE | MAE | RMSE | CVRMSE | MAE | RMSE | CVRMSE |
| **Phoenix** | January | 2 | 34 | 61 | 0.43 | 27 | 58 | 0.4 | 38 | 96 | 0.67 |
| | March | 1 | 76 | 134 | 0.57 | 61 | 121 | 0.51 | 204 | 296 | 0.6 |
| | June | 2 | 42 | 84 | 0.27 | 47 | 87 | 0.28 | 51 | 105 | 0.34 |
| | August | 2 | 35 | 63 | 0.22 | 29 | 58 | 0.2 | 32 | 82 | 0.27 |
| | October | 2 | 24 | 39 | 0.2 | 17 | 35 | 0.18 | 25 | 70 | 0.37 |
| | December | 1 | 41 | 77 | 0.64 | 39 | 76 | 0.63 | 123 | 194 | 1.61 |
| **Chicago** | January | 2 | 22 | 39 | 0.57 | 17 | 35 | 0.51 | 40 | 87 | 1.27 |
| | March | 2 | 33 | 50 | 0.42 | 26 | 47 | 0.4 | 65 | 137 | 1.26 |
| | June | 2 | 58 | 86 | 0.33 | 61 | 98 | 0.38 | 105 | 176 | 0.7 |
| | August | 2 | 54 | 85 | 0.47 | 45 | 77 | 0.43 | 93 | 158 | 0.71 |
| | October | 2 | 32 | 48 | 0.39 | 22 | 41 | 0.33 | 66 | 123 | 1.08 |
| | December | 2 | 16 | 27 | 0.41 | 11 | 22 | 0.34 | 43 | 86 | 1.28 |
| **Miami** | January | 2 | 33 | 55 | 0.35 | 27 | 52 | 0.33 | 54 | 117 | 0.74 |
| | March | 1 | 56 | 86 | 0.35 | 45 | 77 | 0.31 | 47 | 89 | 0.36 |
| | June | 1 | 61 | 93 | 0.39 | 53 | 88 | 0.37 | 57 | 118 | 0.48 |
| | August | 2 | 56 | 90 | 0.4 | 45 | 81 | 0.36 | 71 | 154 | 0.69 |
| | October | 2 | 45 | 80 | 0.5 | 37 | 74 | 0.46 | 50 | 352 | 2.19 |
| | December | 1 | 46 | 84 | 0.7 | 41 | 80 | 0.68 | 50 | 106 | 0.86 |

Table 6: LDMX Models Validation Error Metrics Using TMY Data

| City | Month | Order | All Data | | | Day Time | | | Simple Forecast | | |
|---|---|---|---|---|---|---|---|---|---|---|---|
| | | | MAE | RMSE | CVRMSE | MAE | RMSE | CVRMSE | MAE | RMSE | CVRMSE |
| Phoenix | January | 2 | 44 | 81 | 0.56 | 33 | 69 | 0.48 | 38 | 96 | 0.67 |
| | March | 2 | 81 | 150 | 0.64 | 60 | 126 | 0.53 | 204 | 296 | 0.6 |
| | June | 1 | 47 | 97 | 0.31 | 53 | 97 | 0.31 | 51 | 105 | 0.34 |
| | August | 1 | 48 | 93 | 0.32 | 37 | 78 | 0.27 | 32 | 82 | 0.27 |
| | October | 1 | 30 | 53 | 0.27 | 25 | 51 | 0.26 | 25 | 70 | 0.37 |
| | December | 1 | 54 | 102 | 0.84 | 48 | 95 | 0.79 | 123 | 194 | 1.61 |
| Chicago | January | 2 | 38 | 63 | 0.92 | 24 | 46 | 0.66 | 40 | 87 | 1.27 |
| | March | 1 | 58 | 89 | 0.75 | 37 | 69 | 0.58 | 65 | 137 | 1.26 |
| | June | 1 | 94 | 145 | 0.56 | 79 | 132 | 0.51 | 105 | 176 | 0.7 |
| | August | 1 | 91 | 133 | 0.74 | 58 | 98 | 0.54 | 93 | 158 | 0.71 |
| | October | 1 | 56 | 86 | 0.69 | 33 | 62 | 0.5 | 66 | 123 | 1.08 |

| | December | 1 | 27 | 48 | 0.75 | 17 | 33 | 0.51 | 43 | 86 | 1.28 |
| --- | --- | --- | --- | --- | --- | --- | --- | --- | --- | --- | --- |
| | January | 1 | 52 | 95 | 0.6 | 39 | 79 | 0.5 | 54 | 117 | 0.74 |
| | March | 1 | 55 | 87 | 0.35 | 44 | 77 | 0.31 | 47 | 89 | 0.36 |
| Mimai | June | 1 | 73 | 114 | 0.48 | 58 | 100 | 0.42 | 57 | 118 | 0.48 |
| | August | 1 | 76 | 130 | 0.58 | 58 | 103 | 0.46 | 71 | 154 | 0.69 |
| | October | 1 | 52 | 93 | 0.58 | 42 | 82 | 0.51 | 50 | 352 | 2.19 |
| | December | 1 | 57 | 95 | 0.8 | 46 | 87 | 0.73 | 50 | 106 | 0.86 |

Table 7: Average Error Metrics for LMX and LDMX Models

| City | Data Type | Model | Average CVRMSE | Average MAE |
| --- | --- | --- | --- | --- |
| Phoenix | All 24 hours | LMX | 0.39 | 42 |
| | All 24 hours | LDMX | 0.49 | 51 |
| | Daytime hours | LMX | 0.37 | 37 |
| | Daytime hours | LDMX | 0.44 | 43 |
| | Simple Forecast | | 0.64 | 79 |
| Chicago | All 24 hours | LMX | 0.42 | 36 |
| | All 24 hours | LDMX | 0.69 | 61 |
| | Daytime hours | LMX | 0.4 | 30 |
| | Daytime hours | LDMX | 0.52 | 41 |
| | Simple Forecast | | 1.12 | 69 |
| Miami | All 24 hours | LMX | 0.44 | 50 |
| | All 24 hours | LDMX | 0.55 | 61 |
| | Daytime hours | LMX | 0.42 | 41 |
| | Daytime hours | LDMX | 0.47 | 48 |
| | Simple Forecast | | 0.89 | 55 |

Table 8. Forecast Errors of the LMX and LMX2 Models using TM3 Data

| City | LMX2 | | LMX | | Simple Forecast | |
| --- | --- | --- | --- | --- | --- | --- |
| | RMSE | CVRMSE | RMSE | CVRMSE | RMSE | CVRMSE |
| **Phoenix** | 62 | 0.26 | 86 | 0.44 | 141 | 0.64 |
| **Chicago** | 69 | 0.42 | 73 | 0.55 | 128 | 1.05 |
| **Miami** | 84 | 0.43 | 88 | 0.49 | 156 | 0.89 |

Table 9. Error Metrics for annual and monthly LMX2 Models

| City | LMX2 | | LMX2M | |
| --- | --- | --- | --- | --- |
| | RMSE | CVRMSE | RMSE | CVRMSE |
| Phoenix | 62 | 0.26 | 52 | 0.16 |
| Chicago | 69 | 0.42 | 54 | 0.23 |
| Miami | 84 | 0.43 | 71 | 0.26 |

Table 10. ANN Parameter Tuning Results Using 2009-2013 CPR data for Phoenix

| Parameter | Initial Value | Optimal Value |
|---|---|---|
| Learning Rate | 0.3 | 0.2 |
| Momentum | 0.2 | 0.3 |
| Validation set size | 0 | 0.1 |
| MAE | 35 | 15 |
| RMSE | 54 | 38 |
| CVRMSE | 0.13 | 0.06 |

Table 11. Model Validation Error Metrics for ANN and LMX2 Model using Different Data Sets

| City | Data | ANN | | | LMX2 | | |
|---|---|---|---|---|---|---|---|
| | | MAE | RMSE | CVRMSE | MAE | RMSE | CVRMSE |
| Phoenix | CPR 2009-13 | 15 | 37 | 0.12 | 30 | 52 | 0.16 |
| | CPR 2013 | 16 | 35 | 0.11 | 29 | 45 | 0.14 |
| | TMY 3 | 22 | 47 | 0.15 | 37 | 62 | 0.19 |
| Chicago | CPR 2009-13 | 24 | 48 | 0.19 | 38 | 60 | 0.24 |
| | CPR 2013 | 23 | 44 | 0.19 | 34 | 54 | 0.23 |
| | TMY 3 | 28 | 57 | 0.23 | 42 | 69 | 0.29 |
| Miami | CPR 2009-13 | 35 | 67 | 0.24 | 60 | 93 | 0.34 |
| | CPR 2013 | 34 | 62 | 0.23 | 55 | 82 | 0.3 |
| | TMY 3 | 35 | 64 | 0.23 | 53 | 84 | 0.3 |

**Figures**

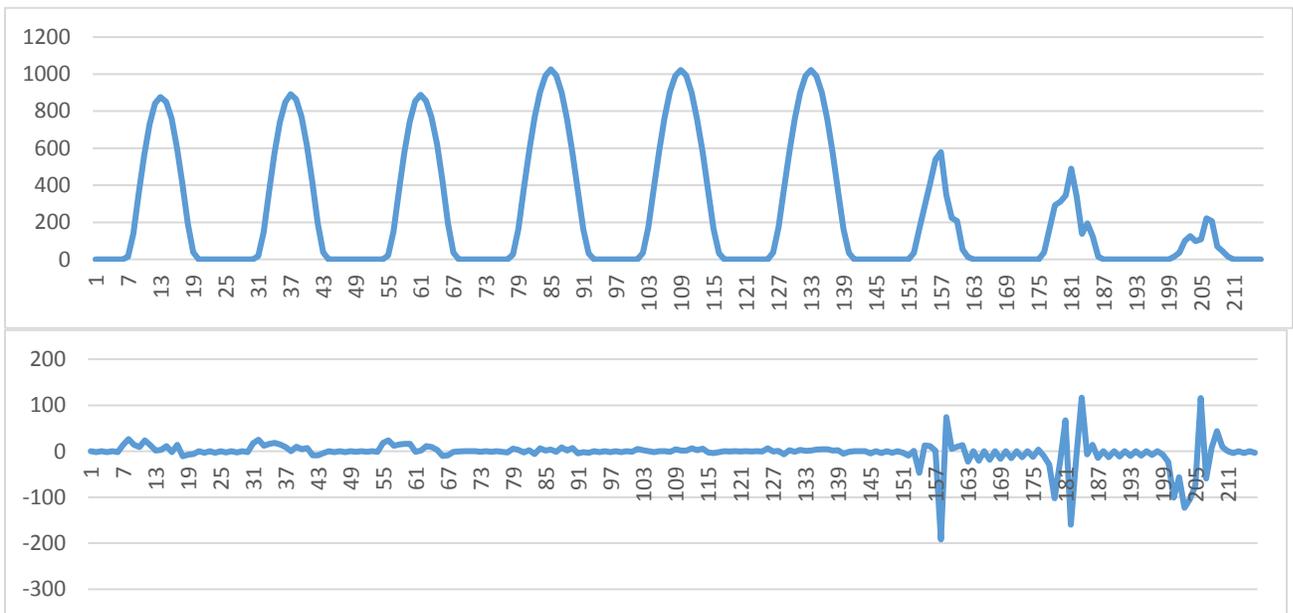

Figure 1: Time series plots over 9-days of (a) solar irradiation and (b) residual plots for SARIMA model to illustrate the how random fluctuations result in very poor models

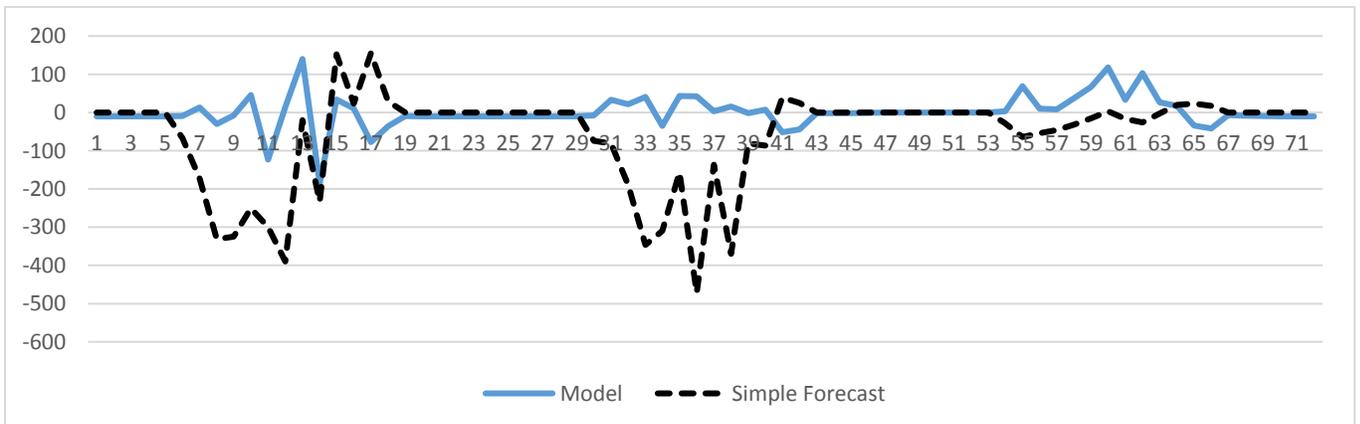

Figure 2. Residual plots of the LMX model and the simple forecast method for three days of the test data set (March, Chicago)

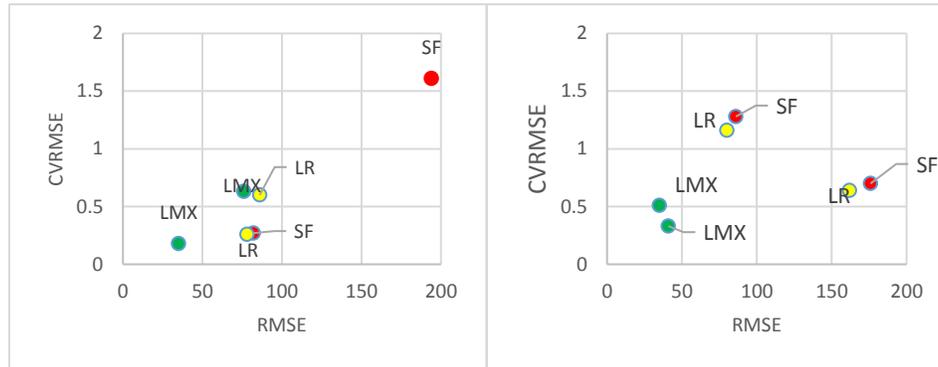

Figure 3: Scatter plot of RMSE vs CVRMSE for Phoenix and Chicago for different models

**References**

[1] Goh, T. N., and K. J. Tan. "Stochastic modeling and forecasting of solar radiation data." *Solar Energy* 19.6 (1977): 755-757.

[2] Mellit, Adel, and Soteris A. Kalogirou. "Artificial intelligence techniques for photovoltaic applications: A review." *Progress in energy and combustion science* 34.5 (2008): 574-632.

[3] Mellit, Adel. "Artificial Intelligence technique for modelling and forecasting of solar radiation data: a review." *International Journal of Artificial intelligence and soft computing* 1.1 (2008): 52-76.

[4] Inman, Rich H., Hugo TC Pedro, and Carlos FM Coimbra. "Solar forecasting methods for renewable energy integration." *Progress in energy and combustion science* 39.6 (2013): 535-576.

[5] Amit Kumar ,Yadav, and S. S. Chandel. "Solar radiation prediction using Artificial Neural Network techniques: A review." *Renewable and Sustainable Energy Reviews* 33 (2014): 772-781.

[6] Qazi, Atika, et al. "The artificial neural network for solar radiation prediction and designing solar systems: a systematic literature review." *Journal of Cleaner Production* 104 (2015): 1-12.

[7] Sfetsos, A., and A. H. Coonick. "Univariate and multivariate forecasting of hourly solar radiation with artificial intelligence techniques." *Solar Energy* 68.2 (2000): 169-178.

[8] Dorvlo, Atsu SS, Joseph A. Jervase, and Ali Al-Lawati. "Solar radiation estimation using artificial neural networks." *Applied Energy* 71.4 (2002): 307-319.

[9] Behrang, M. A., et al. "The potential of different artificial neural network (ANN) techniques in daily global solar radiation modeling based on meteorological data." *Solar Energy* 84.8 (2010): 1468-1480.

[10] Reikard, Gordon. "Predicting solar radiation at high resolutions: A comparison of time series forecasts." *Solar Energy* 83.3 (2009): 342-349.

[11] Benghanem, Mohamed, and Adel Mellit. "Radial basis function network-based prediction of global solar radiation data: application for sizing of a stand-alone photovoltaic system at Al-Madinah, Saudi Arabia." *Energy* 35.9 (2010): 3751-3762.

[12] Rahimikhoob, A., S. M. R. Behbahani, and M. E. Banihabib. "Comparative study of statistical and artificial neural network's methodologies for deriving global solar radiation from NOAA satellite images." *International Journal of Climatology* 33.2 (2013): 480-486.

[13] Ahmad, A., T. N. Anderson, and T. T. Lie. "Hourly global solar irradiation forecasting for New Zealand." *Solar Energy* 122 (2015): 1398-1408.

[14] Lauret, Philippe, et al. "A benchmarking of machine learning techniques for solar radiation forecasting in an insular context." *Solar Energy* 112 (2015): 446-457.

[15] Koca, Ahmet, et al. "Estimation of solar radiation using artificial neural networks with different input parameters for Mediterranean region of Anatolia in Turkey." *Expert Systems with Applications* 38.7 (2011): 8756-8762.

[16] Voyant, Cyril, et al. "Multi-horizon solar radiation forecasting for Mediterranean locations using time series models." *Renewable and Sustainable Energy Reviews* 28 (2013): 44-52.

[17] Wang, Jianzhou, et al. "Forecasting solar radiation using an optimized hybrid model by Cuckoo Search algorithm." *Energy* 81 (2015): 627-644.

[18] Monjoly, Stéphanie, et al. "Hourly forecasting of global solar radiation based on multiscale decomposition methods: A hybrid approach." *Energy* 119 (2017): 288-298.

[19] Ji, Wu, and Keong Chan Chee. "Prediction of hourly solar radiation using a novel hybrid model of ARMA and TDNN." *Solar Energy* 85.5 (2011): 808-817.

[20] Benmouiza, Khalil, and Ali Cheknane. "Forecasting hourly global solar radiation using hybrid k-means and nonlinear autoregressive neural network models." *Energy Conversion and Management* 75 (2013): 561-569.

[21] Voyant, Cyril, et al. "Numerical weather prediction (NWP) and hybrid ARMA/ANN model to predict global radiation." *Energy* 39.1 (2012): 341-355.

[22] Voyant, Cyril, et al. "Hybrid methodology for hourly global radiation forecasting in Mediterranean area." *Renewable Energy* 53 (2013): 1-11.

[23] Voyant, Cyril, et al. "Forecasting method for global radiation time series without training phase: Comparison with other well-known prediction methodologies." *Energy* 120 (2017): 199-208.

[24] Montgomery, Douglas C., Cheryl L. Jennings, and Murat Kulahci. *Introduction to time series analysis and forecasting*. John Wiley & Sons, 2015.

[25] Box, George EP, et al. *Time series analysis: forecasting and control*. John Wiley & Sons, 2015.

[26] Madsen, Henrik. *Time series analysis*. CRC Press, 2007.

[27] Friedman, Jerome, Trevor Hastie, and Robert Tibshirani. The elements of statistical learning. Vol. 1. New York: Springer series in statistics, 2001.

[28] Bishop, Chris M. "Neural networks and their applications." *Review of scientific instruments* 65.6 (1994): 1803-1832.

[29] Haykin, S. "Neural Networks: A Comprehensive Foundation, 1MacMillan." *New York* 19942 (1994): 1.

[30] Hagan, M. T., H. B. Demuth, and M. Beale Neural Network Design. "PWS Publishing Company." *Boston, MA, USA* (1996).

[31] Fausett, Laurene V. *Fundamentals of neural networks*. Prentice-Hall, 1994.

[32] Premalatha, Neelamegam, and Amirtham Valan Arasu. "Prediction of solar radiation for solar systems by using ANN models with different back propagation algorithms." *Journal of applied research and technology* 14.3 (2016): 206-214.